


\font\bigbold=cmbx12
\font\ninerm=cmr9
\font\eightrm=cmr8
\font\sixrm=cmr6
\font\fiverm=cmr5
\font\ninebf=cmbx9
\font\eightbf=cmbx8
\font\sixbf=cmbx6
\font\fivebf=cmbx5
\font\ninei=cmmi9  \skewchar\ninei='177
\font\eighti=cmmi8  \skewchar\eighti='177
\font\sixi=cmmi6    \skewchar\sixi='177
\font\fivei=cmmi5
\font\ninesy=cmsy9 \skewchar\ninesy='60
\font\eightsy=cmsy8 \skewchar\eightsy='60
\font\sixsy=cmsy6   \skewchar\sixsy='60
\font\fivesy=cmsy5
\font\nineit=cmti9
\font\eightit=cmti8
\font\ninesl=cmsl9
\font\eightsl=cmsl8
\font\ninett=cmtt9
\font\eighttt=cmtt8
\font\tenfrak=eufm10
\font\ninefrak=eufm9
\font\eightfrak=eufm8
\font\sevenfrak=eufm7
\font\fivefrak=eufm5
\font\tenbb=msbm10
\font\ninebb=msbm9
\font\eightbb=msbm8
\font\sevenbb=msbm7
\font\fivebb=msbm5
\font\tensmc=cmcsc10


\newfam\bbfam
\textfont\bbfam=\tenbb
\scriptfont\bbfam=\sevenbb
\scriptscriptfont\bbfam=\fivebb
\def\Bbb{\fam\bbfam}

\newfam\frakfam
\textfont\frakfam=\tenfrak
\scriptfont\frakfam=\sevenfrak
\scriptscriptfont\frakfam=\fivefrak
\def\frak{\fam\frakfam}

\def\smc{\tensmc}


\def\eightpoint{%
\textfont0=\eightrm   \scriptfont0=\sixrm
\scriptscriptfont0=\fiverm  \def\rm{\fam0\eightrm}%
\textfont1=\eighti   \scriptfont1=\sixi
\scriptscriptfont1=\fivei  \def\oldstyle{\fam1\eighti}%
\textfont2=\eightsy   \scriptfont2=\sixsy
\scriptscriptfont2=\fivesy
\textfont\itfam=\eightit  \def\it{\fam\itfam\eightit}%
\textfont\slfam=\eightsl  \def\sl{\fam\slfam\eightsl}%
\textfont\ttfam=\eighttt  \def\tt{\fam\ttfam\eighttt}%
\textfont\frakfam=\eightfrak \def\frak{\fam\frakfam\eightfrak}%
\textfont\bbfam=\eightbb  \def\Bbb{\fam\bbfam\eightbb}%
\textfont\bffam=\eightbf   \scriptfont\bffam=\sixbf
\scriptscriptfont\bffam=\fivebf  \def\bf{\fam\bffam\eightbf}%
\abovedisplayskip=9pt plus 2pt minus 6pt
\belowdisplayskip=\abovedisplayskip
\abovedisplayshortskip=0pt plus 2pt
\belowdisplayshortskip=5pt plus2pt minus 3pt
\smallskipamount=2pt plus 1pt minus 1pt
\medskipamount=4pt plus 2pt minus 2pt
\bigskipamount=9pt plus4pt minus 4pt
\setbox\strutbox=\hbox{\vrule height 7pt depth 2pt width 0pt}%
\normalbaselineskip=9pt \normalbaselines
\rm}


\def\ninepoint{%
\textfont0=\ninerm   \scriptfont0=\sixrm
\scriptscriptfont0=\fiverm  \def\rm{\fam0\ninerm}%
\textfont1=\ninei   \scriptfont1=\sixi
\scriptscriptfont1=\fivei  \def\oldstyle{\fam1\ninei}%
\textfont2=\ninesy   \scriptfont2=\sixsy
\scriptscriptfont2=\fivesy
\textfont\itfam=\nineit  \def\it{\fam\itfam\nineit}%
\textfont\slfam=\ninesl  \def\sl{\fam\slfam\ninesl}%
\textfont\ttfam=\ninett  \def\tt{\fam\ttfam\ninett}%
\textfont\frakfam=\ninefrak \def\frak{\fam\frakfam\ninefrak}%
\textfont\bbfam=\ninebb  \def\Bbb{\fam\bbfam\ninebb}%
\textfont\bffam=\ninebf   \scriptfont\bffam=\sixbf
\scriptscriptfont\bffam=\fivebf  \def\bf{\fam\bffam\ninebf}%
\abovedisplayskip=10pt plus 2pt minus 6pt
\belowdisplayskip=\abovedisplayskip
\abovedisplayshortskip=0pt plus 2pt
\belowdisplayshortskip=5pt plus2pt minus 3pt
\smallskipamount=2pt plus 1pt minus 1pt
\medskipamount=4pt plus 2pt minus 2pt
\bigskipamount=10pt plus4pt minus 4pt
\setbox\strutbox=\hbox{\vrule height 7pt depth 2pt width 0pt}%
\normalbaselineskip=10pt \normalbaselines
\rm}


\def\pagewidth#1{\hsize= #1}
\def\pageheight#1{\vsize= #1}
\def\hcorrection#1{\advance\hoffset by #1}
\def\vcorrection#1{\advance\voffset by #1}

\newif\iftitlepage   \titlepagetrue           
\newtoks\titlepagefoot     \titlepagefoot={\hfil} 
\newtoks\otherpagesfoot    \otherpagesfoot={\hfil\tenrm\folio\hfil}
\footline={\iftitlepage\the\titlepagefoot\global\titlepagefalse
           \else\the\otherpagesfoot\fi}

\font\extra=cmss10 scaled \magstep0
\setbox1 = \hbox{{{\extra R}}}
\setbox2 = \hbox{{{\extra I}}}
\setbox3 = \hbox{{{\extra C}}}
\setbox4 = \hbox{{{\extra Z}}}
\setbox5 = \hbox{{{\extra N}}}



\def\ZZZ{{{\extra Z}}\hskip-\wd4\hskip 2.5 true pt{{\extra Z}}}
\def\Zed{\hbox{{\extra\ZZZ}}}       




\def\Z{{\Zed}}

\def\pa{\partial}

\def\tr{{\rm Tr}\,}
\def\pa{\partial}

\def\conf{{\cal Q}}


\def\abstract#1{{\parindent=30pt\narrower\noindent\ninepoint\openup
2pt #1\par}}


\newcount\notenumber\notenumber=1
\def\note#1
{\unskip\footnote{$^{\the\notenumber}$}
{\eightpoint\openup 1pt #1}
\global\advance\notenumber by 1}


\def\frac#1#2{{#1\over#2}}

\def\({\left(}
\def\){\right)}
\def\<{\langle}
\def\>{\rangle}

\def\pmb#1{\setbox0=\hbox{$#1$}%
   \kern-.025em\copy0\kern-\wd0
   \kern.05em\copy0\kern-\wd0
   \kern-.025em\raise.0433em\box0 }


\global\newcount\secno \global\secno=0
\global\newcount\meqno \global\meqno=1
\global\newcount\appno \global\appno=0
\newwrite\eqmac
\def\romappno{\ifcase\appno\or A\or B\or C\or D\or E\or F\or G\or H
\or I\or J\or K\or L\or M\or N\or O\or P\or Q\or R\or S\or T\or U\or
V\or W\or X\or Y\or Z\fi}
\def\eqn#1{
        \ifnum\secno>0
        \eqno(\the\secno.\the\meqno)\xdef#1{\the\secno.\the\meqno}
          \else\ifnum\appno>0
  \eqno({\rm\romappno}.\the\meqno)\xdef#1{{\rm\romappno}.\the\meqno}
          \else
            \eqno(\the\meqno)\xdef#1{\the\meqno}
          \fi
        \fi
\global\advance\meqno by1 }


\global\newcount\refno
\global\refno=1 \newwrite\reffile
\newwrite\refmac
\newlinechar=`\^^J
\def\ref#1#2{\the\refno\nref#1{#2}}
\def\nref#1#2{\xdef#1{\the\refno}
\ifnum\refno=1\immediate\openout\reffile=refs.tmp\fi
\immediate\write\reffile{
     \noexpand\item{[\noexpand#1]\ }#2\noexpand\nobreak}
     \immediate\write\refmac{\def\noexpand#1{\the\refno}}
   \global\advance\refno by1}
\def\semi{;\hfil\noexpand\break ^^J}
\def\nl{\hfil\noexpand\break ^^J}
\def\refn#1#2{\nref#1{#2}}

\def\ann#1#2#3{{\sl Ann. Phys.} {\bf {#1}} (19{#2}) #3}
\def\cmp#1#2#3{{\sl Commun. Math. Phys.} {\bf {#1}} (19{#2}) #3}

\def\jmp#1#2#3{{\sl J. Math. Phys.} {\bf {#1}} (19{#2}) #3}

\def\mplA#1#2#3{{\sl Mod.  Phys.  Lett.} {\bf A{#1}} (19{#2}) #3} 
 
\def\np#1#2#3{{\sl Nucl.  Phys.} {\bf B{#1}} (19{#2}) #3} 

\def\plB#1#2#3{{\sl Phys.  Lett.} {\bf {#1}B} (19{#2}) #3}

\def\prD#1#2#3{{\sl Phys.  Rev.} {\bf D{#1}} (19{#2}) #3} 
 
\def\prp#1#2#3{{\sl Phys.  Rep.} {\bf {#1}C} (19{#2}) #3}

\refn\DDDPW
{
A.\ D'Adda, A.C.\ Davis, P.\ Di Vecchia
\plB{121}{83}{335}; \nl
A.M.\ Polyakov, P.B.\ Wiegmann,
\plB{141}{84}{223}.
}
\refn\W
{E.\ Witten,
\cmp{92}{84}{455}.
}
\refn\GK
{See, {\it e.g.}, 
K.\ Gawedzki, A.\ Kupiainen,
\np{320}{89}{625} and references therein.
}
\refn\Wi
{E.\ Witten,
\prD{44}{91}{314}.
}
\refn\FORTWTF
{L.\ Feh\'er, L.\ O'Raifeartaigh, 
 P.\ Ruelle, I.\ Tsutsui, A.\ Wipf,
\prp{222}{92}{1}; \nl
I.\ Tsutsui, L.\ Feh\'er,
\plB{294}{92}{209}.
}
\refn\ORWFR
{
E.\ Ogievetskii, N.\ Reshetikhin, P.\ Wiegmann,
\np{280}{87}{45}; \nl
L.D.\ Faddeev, N.Yu.\ Reshetikhin,
\ann{167}{86}{227}.
}
\refn\MS
{P.\ Mejean, F.A.\ Smirnov,
{\it Form-Factors for Principal Chiral Field
Model with Wess-Zumino-Novikov-Witten Term}, 
hep-th/9609068.
}
\refn\KZ
{V.G.\ Knizhnik, A.B.\ Zamolodchikov,
\np{247}{84}{83}.
}
\refn\DS
{
C.\ Devchand, J.\ Schiff,
{\it Hidden Symmetries of the 
Principal Chiral Model Unveiled},
hep-th/9611081.
}
\refn\ZZ
{
A.B.\ Zamolodchikov, Al.B.\ Zamolodchikov,
\np{379}{92}{602} and references threrein.
}
\refn\M
{G.\ Mackey,
\lq\lq Induced Representation of Groups and Quantum 
Mechanics\rq\rq,
Benjamin, New York, 1969.
}
\refn\I
{C.J.\ Isham,
{\it in} \lq\lq Relativity, Groups, and Topology\rq\rq 
$\,$ (B.S.\ DeWitt and R.\ Stora, Eds.), North-Holland, 
Amsterdam, 1984.
}
\refn\LL
{N.P.\ Landsman, N.\ Linden, 
\np{365}{91}{121}.
}
\refn\OKMT
{
Y.\ Ohnuki, S.\ Kitakado, 
\mplA{7}{92}{2477}; 
\jmp{34}{93}{2827}; \nl
D.\ McMullan, I.\ Tsutsui, 
\plB{320}{94}{287};
\ann{237}{95}{269}; \nl
S.\ Tanimura, I.\ Tsutsui,
{\it Inequivalent Quantizations and
Holonomy Factor from the Path-Integral Approach}, 
hep-th/9609089, to appear in {\sl Ann. Phys.}
}
\refn\Wit
{E.\ Witten,
\np{223}{83}{422}.
}
\refn\KTT
{H.\ Kobayashi, S.\ Tanimura, I.\ Tsutsui,
{\it 
Quantum Mechanically Induced Hopf Term
in the $O(3)$ Nonlinear Sigma Model}, hep-th/9705183.
}
\refn\WZ
{Y.S.\ Wu, A.\ Zee,
\np{272}{86}{322}.
}


\pageheight{23cm}
\pagewidth{15cm}
\hcorrection{-1mm}
\magnification= \magstep1
\def\bsk{%
\baselineskip= 16pt plus 1pt minus 1pt}
\parskip=5pt plus 1pt minus 1pt
\tolerance 8000



\null

\rightline{KEK Preprint 97-84}
\rightline{June 1997}

\smallskip
\vfill
{\baselineskip=18pt

\centerline{\bigbold 
Quantum Mechanically Induced Wess-Zumino Term}
\centerline{\bigbold 
in the Principal Chiral Model}

\vskip 30pt

\centerline{
\smc 
Hitoshi Miyazaki
\quad {\rm and} \quad 
Izumi Tsutsui\note
{E-mail:\quad tsutsui@tanashi.kek.jp}
}

\vskip 25pt

{
\baselineskip=13pt
\centerline{\it 
Institute of Particle and Nuclear Studies}
\centerline{\it 
High Energy Accelerator Research Organization (KEK),
Tanashi Branch}
\centerline{\it Midori, Tanashi, Tokyo 188} 
\centerline{\it Japan}
}

\vskip 80pt

\abstract{%
{\bf Abstract.}\quad
It is argued that, in the two dimensional principal
chiral model, 
the Wess-Zumino term
can be induced quantum mechanically, allowing 
the model with the critical value of the coupling constant
$\lambda^2 = 8\pi/\vert k \vert$ 
to turn into the Wess-Zumino-Novikov-Witten model 
at the quantum level.
The Wess-Zumino term emerges from
the inequivalent quantizations possible on a sphere 
hidden in the configuration space of the original model.
It is shown that the Dirac monopole potential, which is induced
on the sphere in the inequivalent quantizations, 
turns out to be 
the Wess-Zumino term in the entire configuration space.
}

\vfill\eject

\bsk


\secno=1 \meqno=1 

\noindent
{\bf 1. Introduction}
\medskip

The Wess-Zumino-Novikov-Witten (WZNW) model defined 
by the action\note{
Convention: We use $(x^0,x^1) = (t,x)$ for spacetime coordinates,
$k \in \Z$ for the level of the Kac-Moody algebra,
$\{T_a = \sigma_a/2i; a = 1, 2, 3\}$ for a basis
of the Lie algebra of $SU(2)$, and the normalized trace
$\tr(T_aT_b) = \delta_{ab}$ given by 
$\tr := (-2)$ times the matrix trace.
}
$$
I(g)
      = {1 \over{2 \lambda^2}} \int_{S^2} d^2x\, 
  \tr (g^{-1} \pa_\mu g)^2
                 - {k\over{24 \pi}} \int_{D^3} 
\tr (g^{-1}dg)^3 
\eqn\wzaction
$$
at the coupling constant $\lambda^2 = 8\pi/\vert k \vert$ is
perhaps one of the most useful field theory models
in two dimensions.  It was originally considered for 
constructing 
effective actions in non-Abelian theories [\DDDPW] 
and further attracted attention
as the model for non-Abelian bosonization [\W],
but it was soon realized that the model 
could also be used for 
a variety of 
purposes, such as to furnish models of 
rational conformal field
theories [\GK], to find black hole solutions [\Wi]
and to construct integrable models by Hamiltonian
reduction [\FORTWTF], to mention a few.
The key ingredient of the WZNW model is the 
second term in (\wzaction), {\it i.e.},
the Wess-Zumino term, which is defined on a three
dimensional disc $D^3$ whose boundary is the spacetime
$S^2$.  The addition of this term to the first term 
in (\wzaction) --- the first term alone 
gives the action of another useful 
integrable model, the  
principal chiral model [\ORWFR] (see also [\MS]) ---
has major
ramifications in the physics of the original
model bestowing on
it the Kac-Moody algebra and hence conformal symmetry [\KZ].
(It is, however, argued that the principal chiral model 
also possesses a symmetry algebra of infinite dimensions [\DS],
and that the addition of the Wess-Zumino term still preserves
the integrability of the model [\ZZ].)
The aim of the present note is to point out the possibility
that this Wess-Zumino term may be induced quantum mechanically,
once we accept the outcome of quantum mechanics on a coset space 
and apply it to the principal chiral model.

The basic reason for the emergence of the Wess-Zumino 
term is
that there exists a sector represented by
a sphere $S^2$ in the configuration space
of the principal chiral model, and quantization on
this sector gives rise to a
quantum effect in the form of the Wess-Zumino term
which modifies the classical action of the model,
just as the $\theta$-term in QCD which 
arises from the possible 
inequivalent quantizations (superselection
sectors) modifies the classical action.
In fact, it has been known that
quantum mechanics on a sphere admits inequivalent 
quantizations [\M,\I] labelled by
an integer.  The point is that these inequivalent 
quantizations come equipped with the induced potential
of the Dirac monopole 
[\LL] (see also [\OKMT]), which
is the potential mentioned by Witten
[\Wit] as a prototype of the Wess-Zumino term.
Here we shall show that
it is more than a prototype: the Wess-Zumino 
term does reduce to 
the potential term
of the Dirac monopole in a certain limit --- a fact indicating 
that the Wess-Zumino term
is induced quantum mechanically.  For simplicity we restrict
ourselves to the case of the group $SU(2)$, but our discussion
remains essentially 
the same for more general groups for which the
Wess-Zumino term can be defined.

\secno=2 \meqno=1 

\bigskip
\noindent
{\bf 2. Sphere in the configuration space}
\medskip

We begin by extracting the degrees of freedom
of a sphere $S^2$ from the
configuration space $\conf$ based on the
decomposition of the field with respect to the
homotopy group $\pi_2(\conf)$.
The configuration space $\conf$ 
of the principal chiral model of interest is the space  
of $SU(2)$-valued fields on a circle $S^1$ taking a fixed
value at a fixed point.  
More explicitly, if we let $x \in [0, 2\pi]$ 
be the coordinate on the circle $S^1$, then the space $\conf$ 
consists of fields $g(x) \in SU(2)$ which satisfy
the periodic boundary condition $g(2\pi) = g(0)$ 
with a fixed value, say, $g(0) = 1$.  In other words,
$\conf$ is defined to be
the space of based maps from $S^1$ to $SU(2)$:
$$
\conf = {\rm Map}_0(S^1, SU(2)).
\eqn\cfcd
$$

The important property of the space of based maps 
is that the homotopy groups of the space 
can be related to that of the target
space $SU(2) \simeq S^3$ (see, {\it e.g.}, [\KTT]),
$$
\pi_k(\conf) = \pi_{k+1}(S^3), 
\qquad \hbox{for} \quad k = 0, 1, 2,
\ldots.
\eqn\no
$$
In particular, we have
$$
\pi_0(\conf) = \pi_{1}(S^3) = 0, \qquad
\pi_1(\conf) = \pi_{2}(S^3) = 0,
\eqn\no
$$
which shows that the configuration space $\conf$ is  
path-connected and simply-connected.
On the other hand, we also have 
$$
\pi_2(\conf) = \pi_{3}(S^3) = \Z,
\eqn\homotopy
$$
which implies that there exists a sphere in $\conf$
which cannot shrink smoothly to a point, that is,
there exists a hole in $\conf$ obstructing such a smooth
process.

As is well known, the integral expression that assigns
the integers $n \in \Z$ of the homotopy group (\homotopy)
to a map 
$g: S^3 \rightarrow SU(2)$ is the winding number formula, 
$$
w(g) := {1\over{48\pi^2}} \int_{S^3}\, \tr (g^{-1}dg)^3,
\eqn\winding
$$
where 
it is understood that $\tr (g^{-1}dg)^3$ in (\winding) 
is the pullback of the 3-form on the $SU(2)$ manifold
onto the parameter space $S^3$.
To provide the parameter space $S^3$,
we consider
a two-parameter family of configurations $g(x;t,s)$
by introducing a set of parameters 
$(t, s) \in \Sigma^2 := [0, T] \times [0, 1]$.
If we combine the set with the space parameter $x$, we have
a map $g: \Sigma^3 \rightarrow SU(2)$ 
with $\Sigma^3 := [0, 2\pi] \times \Sigma^2$, but this
can be regarded as a map
$g: S^3 \rightarrow SU(2)$ if it takes the fixed value on 
the boundary:
$$
g(x;t,s) = 1 \qquad \hbox{for} \quad (x,t,s) \in \pa\,\Sigma^3.
\eqn\bc
$$
Note that, once the map satisfies the boundary
condition (\bc), then it also 
gives a map $g: S^2 \rightarrow SU(2)$
at a constant slice of any of 
the three parameters.
Below we identify our spacetime with
the $S^2$ obtained at, say, $s = 1/2$, 
providing the time evolution of the 
configuration by $g(x,t) := g(x;t,{1\over 2})$.

We now construct explicitly a configuration 
satisfying 
the boundary condition (\bc) and possessing the winding number $n$.
Let $H = U(1)$ be the subgroup of
$SU(2)$ generated by $T_3$, and 
consider the maps,
$\sigma: \Sigma^2 \rightarrow SU(2)$ and 
$\tau: [0, 2\pi] \rightarrow H$, given by
$$
\sigma_n(t, s) = e^{ \pi s T_1 }\,e^{ 2n\pi (t/T) T_3 }\,
e^{ - \pi s T_1 },
\eqn\smap
$$
and
$$
\tau(x) = e^{x T_3}.
\eqn\tmap
$$
At the boundary $\pa\,\Sigma^3$ they take the values,
$$
\eqalign{
\sigma_n(t,0) &= e^{ 2n\pi (t/T) T_3 }, \cr 
\sigma_n(0,s) &= 1, \cr
}
\qquad
\eqalign{
\sigma_n(t,1) &= e^{ - 2n\pi (t/T) T_3 }, \cr
\sigma_n(T,s) &= (-1)^n, \cr
}
\eqn\sbc
$$
and
$$
\tau(0) = 1, \qquad \tau(2\pi) = -1.
\eqn\tbc
$$
Then, it follows from (\sbc) and (\tbc) 
that the configuration,
$$
g_n(x;t,s) 
:= \sigma_n(t,s)\, \tau^{-1}(x)\, \sigma^{-1}_n(t,s) \, \tau(x),
\eqn\nwind
$$
{}fulfills the boundary condition (\bc).  
That the configuration $g_n$ in 
(\nwind) has indeed the winding number $n$
can be confirmed by a straightforward calculation, 
$$
w(g_n) = {{(-3)\!\cdot\!(-2\pi)\!\cdot\!2}\over {48\pi^2}} 
\int_{\pa\,\Sigma^2} \tr T_3 (\sigma_n^{-1} d\sigma_n)
= n.
\eqn\wnumber
$$

More generally, we may consider any map
$\sigma_n: \Sigma^2 \rightarrow SU(2)$  
which becomes $H$-valued on the boundary,
$$
\sigma_n(t, s) \in H 
\qquad \hbox{for} \quad (t,s) \in 
\pa\,\Sigma^2.
\eqn\sbc
$$
In fact, this condition is all we need
to ensure that the configuration $g_n(x;t,s)$, given
by (\nwind) with the help of the map $\tau(x)$ in (\smap),
satisfies the boundary
condition (\bc).  Moreover, the condition (\sbc) implies that
on the boundary $\pa\,\Sigma^2$ the map 
$\sigma_n$ furnishes a map $S^1 \rightarrow H = U(1)$ for which
the winding number related to $\pi_1(U(1)) = \Z$ can be assigned
by the formula,
$$
Q(\sigma_n) :=  
-{1\over{4\pi}}\int_{\pa\,\Sigma^2} \tr T_3 
(\sigma_n^{-1} d\sigma_n).
\eqn\wnsoliton
$$
Hence we find from (\wnumber) that for 
$g_n$ of the type (\nwind) we just have
$w(g_n) = - Q(\sigma_n)$, and hence by choosing the $\sigma_n$
so that $Q(\sigma_n) = - n$ we obtain the configuration $g_n$
possessing the winding number $n$.

Observe that 
the configuration $g_n$ in 
(\nwind) remains unchanged under the $H = U(1)$ 
\lq gauge transformation',
$$
\sigma(t,s) \longrightarrow \sigma(t,s)\,h(t,s) 
\qquad \hbox{for} \quad h(t,s) \in H. 
\eqn\gaugetrsf
$$
Note that for $h(t,s)$ to be well-defined over $\Sigma^2$, 
it must become a trivial map
when restricted to the boundary $\pa\,\Sigma^2$, that is,
$Q(h) = 0$.  Accordingly, the winding number (\wnsoliton)
assigned to $\sigma_n$ is gauge invariant.

The redundancy under the gauge transformation 
(\gaugetrsf) suggests that we should consider
the \lq physical target space' of the map $\sigma_n$ 
which contributes to 
the configuration $g_n$.  The physical space is
given by quotienting $\sigma_n$ with respect to  
gauge symmetry under the $U(1)$ transformation (\gaugetrsf), 
that is, by $SU(2)/U(1) \simeq S^2$.   
This space may be explicitly obtained by the 
canonical projection ${\rm pr}: SU(2) \rightarrow S^2$ given
by the adjoint action on the element $T_3$:
$$
{\rm pr}(\sigma_n) := \sigma_n \, T_3\, \sigma_n^{-1} 
= q_1 T_1 + q_2 T_2 + q_3 T_3.
\eqn\projection
$$
The vector ${\bf q} = (q_1,q_2,q_3)$ 
is then subject to 
the constraint ${\bf q}^2 
= \tr(\sigma_n \, T_3\, \sigma_n^{-1})^2 = 1$, showing that
the target space is $S^2$.  In particular, 
if we parametrize $\sigma_n$ as
$$
\sigma_n(t, s) = e^{\alpha(t,s) T_3}\,e^{\beta(t,s)T_2 }\,
e^{\gamma(t,s)T_3},
\eqn\gens
$$
with $\alpha(t,s)$, $\beta(t,s)$ and $\gamma(t,s)$ 
being some functions 
on $\Sigma^2$ which respect 
the boundary condition (\sbc), then the canonical
projection furnishes the polar 
coordinates 
${\bf q} = (\sin\beta\cos\alpha, 
\sin\beta\sin\alpha, \cos\beta)$ on $S^2$ which are
gauge invariant.

Now the point is that, given any map
$g: S^3 \rightarrow SU(2)$ (not just those of the 
particular type (\nwind)) 
whose winding number is $n$, we can decompose it as
$$
g(x;t,s) = g_n(x;t,s)\,\hat g(x;t,s),
\eqn\decomp
$$
where  
$\hat g$ is a map $S^3 \rightarrow SU(2)$ with vanishing
winding number $w(\hat g) = 0$.  The proof is simple --- we
just consider the product $g_n^{-1}(x;t,s)\, g(x;t,s)$ 
and evaluate its winding number.  Thanks to the additivity
of the winding number under multiplication,
$$
w(g\,g') = w(g) + w(g'),
\eqn\additivity
$$
which holds for any $g$, $g': S^3 \rightarrow SU(2)$,
we find $w(g_n^{-1}g) = 0$.  Then the decomposition
(\decomp) follows if we write the product 
as $\hat g := g_n^{-1}g$.
We therefore see that
the configuration $g \in \conf$ appearing at any fixed point
of the parameters $(t,s) \in \Sigma^2$ 
may be factorized in the form
(\decomp) in which the degrees of freedom of a sphere
$S^2$ is isolated in $g_n$ keeping all other (infinite)
degrees of freedom of $g$ in $\hat g$.

\secno=3 \meqno=1 

\bigskip
\noindent
{\bf 3. Induced Wess-Zumino term}
\medskip

Having extracted the $S^2$ degrees of freedom from
the configuration space $\conf$, we next consider
quantization of the principal chiral model on
the $SU(2)$ group manifold which is 
governed by the action
(\wzaction) without the Wess-Zumino term.
Since our configuration space $\conf$ is 
nowhere close in structure to the Euclidean
space (in view of (\homotopy), for instance),
we have to develop in principle a proper
quantization framework applicable to the space $\conf$ 
other than the canonical quantization scheme 
which works only for spaces isomorphic
to the Euclidean space.
Since such a framework has not yet been developed,
here we shall be content with the latter scheme
amending it by the quantum
effects which arise 
due to the fact that the space $\conf$ is not
Euclidean.  In particular, we will be interested in
the quantum effect which emerges when quantizing
the $S^2$ degrees of freedom in $\conf$.

The quantum effect that arises on a sphere $S^2$ has been
known [\LL] and is given by an induced potential of the
type of the Dirac monopole.  To be more explicit, 
if we use the polar coordinates for the sphere $S^2$,
the induced action reads
$$
I_{\rm ind} = \int_0^T  {\bf A}
\!\cdot\! \dot{\bf q}\, dt 
= \int_0^T \, {n \over 2}(1 - \cos\beta)\,
\dot{\alpha}\, dt,
\eqn\indaction
$$
where $n/2 \in \Z/2$ is the (quantized) magnetic
charge (${n\hbar c}/2e$ in the standard unit) which
characterizes the inequivalent quantization or
the superselection sector we are in.  
In the following we argue that the induced action (\indaction)
arises in the principal chiral model in the guise of 
the Wess-Zumino term, 
$$
\Gamma(g) := - {1 \over{24 \pi}} \int_{D^3} 
\tr (g^{-1}dg)^3,
\eqn\wzterm
$$
with the coefficient $k$ being (twice) the monopole charge.
We do this by showing that the Wess-Zumino
term does contain the induced term (\indaction)
and that it is the only term of that property which is 
local and Lorentz invariant.  

To this end, we first note that
by means of the gauge 
transformation (\gaugetrsf)
we can transform $\sigma_n$ in such a way that the new
$\sigma_n$ becomes constant 
over the boundary $\pa\,\Sigma^2$ except for
the edge $(t,s = 1)$ with $t \in [0,T]$.  (We cannot render
$\sigma_n$ constant everywhere on $\pa\,\Sigma^2$ since 
the gauge transformation cannot alter the winding number.) 
This allows us to regard the map $\sigma_n : \Sigma^2 
\rightarrow S^2$ as a map $\sigma_n : D^2 
\rightarrow S^2$ where the boundary of the disc $D^2$ 
consists of the edge at $s = 1$ mentioned above.

Recall that, in terms of the parameters we are
using, 
the disc $D^3$ where the Wess-Zumino term (\wzterm) 
is defined corresponds to the domain
$[0,2\pi] \times [0,T] \times [0,{1\over 2}]$.  
Denoting $\widetilde\Sigma^2
:= [0,T] \times [0,{1\over 2}]$, we see that
the new $\sigma_n$ can still be regarded as a map
$\sigma_n : D^2 \rightarrow S^2$ even when restricted
to the domain $\widetilde\Sigma^2$ where now the boundary
$\pa D^2 \simeq S^1$ consists of the edge at $s = 1/2$.
With this in mind, we evaluate 
the Wess-Zumino term using the decomposition (\decomp) as
$$
\Gamma(g) = \Gamma(g_n) + \Gamma(\hat g) + 
{1\over{8 \pi}}\int_{S^2}\tr(g_n^{-1}dg_n)(d\hat g \hat g^{-1}).  
\eqn\wzall
$$
Then, the substitution of (\nwind) and (\gens)
into $\Gamma(g_n)$ yields
$$
\Gamma(g_n) = -2\pi\!\cdot\!{1 \over{4\pi}} 
  \int_{\pa\,\widetilde\Sigma^2} 
  \tr T_3 (\sigma_n^{-1} d\sigma_n)
= {1 \over 2}
  \int_0^T \,(1 - \cos\beta)\,\dot{\alpha}\, dt,
\eqn\wzdirac
$$
where we have used $\alpha(t) := \alpha(t, {1\over 2})$ and 
$\beta(t) := \beta(t, {1\over 2})$.
Comparing this with (\indaction), we see that,
modulo
the two extra terms involving $\hat g$ in (\wzall),
the second term $k \Gamma(g)$ in the action
(\wzaction) reduces precisely to
the induced Dirac potential term (\indaction) with
the magnetic charge $k/2$.

Let us now examine the possibility  
whether any local and Lorentz invariant term 
$\Delta I(g)$
can be added to the Wess-Zumino term $\Gamma(g)$
preserving the property that the action reduces to the
Dirac potential term when $g$ becomes $g_n$.
This is equivalent to finding a local and Lorentz 
invariant term $\Delta I(g)$ for 
which $\Delta I(g_n) = 0$, but this is impossible
since $g_n$, being dependent only on time (at $s = 1/2$), 
cannot vanish by any Lorentz invariant operations
(differentiation and/or multiplication) available.
We therefore conclude that the Wess-Zumino
term is induced quantum mechanically upon
quantizing the principal chiral model and that,
in particular for
$\lambda^2 = 8\pi/\vert k \vert$ 
with $k$ some integer, we
obtain the WZNW model at the quantum level.

{}Finally, we mention that the Wess-Zumino term
(in four dimensions) 
has previously been found to   
provide the Dirac monopole potential in
the sense of the functional potential [\WZ], by
identifying the term with the functional 
holonomy factor along the closed
loop $C$ in $\conf$ formed by the evolution 
of the configuration during the time 
interval $[0,T]$.  According to this prescription, 
the functional potential ${\cal A}$ in two 
dimensions can be found by writing 
$k\Gamma(g) = \int_C {\cal A}$, and from this 
the functional curvature reads
$$
{\cal F} = \delta{\cal A} = - {k \over{24 \pi}} \int_{S^1} 
\tr [(g^{-1}dg)(g^{-1}\delta g)^2],
\eqn\curvature
$$
where $\delta$ stands for the functional exterior derivative
and the $S^1$ is the integral domain over 
the space interval $[0,2\pi]$.  Take, then, a sphere $S^2$
in $\conf$ which contains the loop $C$ on it 
and evaluate the flux penetrating the sphere,
$\int_{S^2} {\cal F}$.  Obviously, we may use the set of
parameters $(t,s)$ for the purpose of furnishing            
the sphere in $\conf$ if we preserve the boundary condition 
(\bc).  This allows us to merge the sphere $S^2$ and 
the space $S^1$ to form $S^3$ and thereby obtain
$\int_{S^2} {\cal F} = 
- 2\pi k\, w(g)$, which shows that the flux is 
quantized just as the flux of the usual Dirac
monopole potential.  We, however, stress that, despite
this similarity, the functional monopole 
potential is distinct from 
the induced but conventional monopole 
potential (\wzdirac) discussed above.

\vfill\eject

\baselineskip= 13.5pt

  \vfill\eject\immediate\closeout\reffile
  \centerline{{\bf References}}\bigskip\frenchspacing%
  \input refs.tmp\vfill\eject\nonfrenchspacing

\bye